\documentclass[aps,prl,twocolumn,superscriptaddress]{revtex4}
\usepackage{amssymb}
\usepackage{amsmath}
\usepackage{epsfig}
\usepackage{color}
\usepackage{graphics, graphicx}
\usepackage{bbold}
\usepackage{psfrag}
\usepackage{mathcomp}
\usepackage{subfigure}
\usepackage{verbatim}
\usepackage{color}
\usepackage[colorlinks,citecolor=blue]{hyperref}

\begin{document}

\title{Symmetry-protected topological states for interacting fermions in alkaline-earth-like atoms}
\author{Xiaofan Zhou}
\affiliation{State Key Laboratory of Quantum Optics and Quantum Optics Devices, Institute
of Laser spectroscopy, Shanxi University, Taiyuan 030006, China}
\affiliation{Collaborative Innovation Center of Extreme Optics, Shanxi University, Taiyuan, Shanxi 030006, China}
\author{Jian-Song Pan}
\affiliation{Key Laboratory of Quantum Information, University of Science and Technology of China, CAS, Hefei, Anhui, 230026, China}
\affiliation{Synergetic Innovation Center of Quantum Information and Quantum Physics, University of Science and Technology of China, Hefei, Anhui 230026, China}
\author{Zheng-Xin Liu}
\email{liuzxphys@ruc.edu.cn}
\affiliation{Department of Physics, Renmin University of China, Beijing 100872, China}
\affiliation{Beijing Key Laboratory of Opto-electronic Functional Materials and Micro-nano Devices,
Renmin University of China, Beijing 100872, China}
\author{Wei Zhang}
\email{wzhangl@ruc.edu.cn}
\affiliation{Department of Physics, Renmin University of China, Beijing 100872, China}
\affiliation{Beijing Key Laboratory of Opto-electronic Functional Materials and Micro-nano Devices,
Renmin University of China, Beijing 100872, China}
\author{Wei Yi}
\email{wyiz@ustc.edu.cn}
\affiliation{Key Laboratory of Quantum Information, University of Science and Technology of China, CAS, Hefei, Anhui, 230026, China}
\affiliation{Synergetic Innovation Center of Quantum Information and Quantum Physics, University of Science and Technology of China, Hefei, Anhui 230026, China}
\author{Gang Chen}
\email{chengang971@163.com}
\affiliation{State Key Laboratory of Quantum Optics and Quantum Optics Devices, Institute
of Laser spectroscopy, Shanxi University, Taiyuan 030006, China}
\affiliation{Collaborative Innovation Center of Extreme Optics, Shanxi University, Taiyuan, Shanxi 030006, China}
\author{Suotang Jia}
\affiliation{State Key Laboratory of Quantum Optics and Quantum Optics Devices, Institute
of Laser spectroscopy, Shanxi University, Taiyuan 030006, China}
\affiliation{Collaborative Innovation Center of Extreme Optics, Shanxi University, Taiyuan, Shanxi 030006, China}
\date{\today }

\begin{abstract}
We discuss the quantum simulation of symmetry-protected topological (SPT) states for interacting fermions in quasi-one-dimensional gases of alkaline-earth-like atoms such as $^{173}$Yb. Taking advantage of the separation of orbital and nuclear-spin degrees of freedom in these atoms, we consider Raman-assisted spin-orbit couplings in the clock states, which, together with the spin-exchange interactions in the clock-state manifolds, give rise to SPT states for interacting fermions. We numerically investigate the phase diagram of the system, and study the phase transitions between the SPT phase and the symmetry-breaking phases. The interaction-driven topological phase transition can be probed by measuring local density distribution of the topological edge modes.
\end{abstract}

\maketitle

\emph{Introduction}.--
The study of symmetry-protected topological (SPT) phases has significantly improved our understanding of topological matters~\cite{guwen2009,pollmann2010}. In contrast to intrinsic topological orders with long-range entanglements~\cite{wen89,WenNiu90,Wen90}, SPT phases feature short-range-entangled ground states with bulk gaps, and can have gapless/degenerate edge excitations as long as the protecting symmetries are not broken. Notable examples of SPT states range from the Haldane phase in interacting spin chains~\cite{Haldane83}, which features bosonic edge modes (bosonic SPT state); to topological insulators in free fermions~\cite{hall1,hall2,hall3,hall4,hall5}, whose edge modes are fermionic (fermionic SPT state). In recent years, both the bosonic SPT states and the non-interacting fermionic SPT states have been well classified~\cite{ChenGuWen, ChenGuLiuWen, Kitaev, Shinsy}. The study of interacting SPT phases with fermionic edge modes, however, is still in progress~\cite{Kitaev2,GuWen,WangSenthil,HQWu}. In particular, an experimentally accessible system capable of stabilizing interacting fermionic SPT phases is still lacking.

In this work, we discuss the quantum simulation of SPT states for interacting fermions using alkaline-earth-like atoms. With two valence electrons, these atoms feature long-lived excited states and fermionic isotopes with non-zero nuclear spins. The nuclear- and the orbital-degrees of freedom are decoupled in the ground $^1S_0$ (the so-called $|g\rangle$ orbital) and the meta-stable excited $^3P_0$ (the $|e\rangle$ orbital) manifolds, which enables flexible control of these so-called clock states. While the high level of quantum control has led to numerous applications in quantum metrology, quantum information and quantum simulation using the clock states~\cite{AE1,AE2,AEnew7,AE3,AEnew3,AE4,congjun03,AE5,AEnew1,AEnew4,AEnew5,AEnew2,phasePRL,phaseEPL,quella13,ofr1,ofr2,ofr3,AEnew8,phases1,quella15,AEnew9,phases2,phases3,ye2016old,ye2016,fallani2016,gyuboong}, the recently discovered orbital Feshbach resonance in $^{173}$Yb atoms further enriches the available toolbox, offering exciting possibilities of studying strongly interacting fermionic systems using these atoms~\cite{ren1,ofrexp1,ofrexp2}.

Taking advantage of these features, we show that a topological phase of interacting fermions can be realized in a quasi-one-dimensional (1D) cold gas of alkaline-earth-like atoms. Such a topological phase is protected by the $U(1)$ particle-number-conservation and the chiral symmetries, which form an anti-unitary group $U(1)\times Z_2^T$. The SPT phase has fermionic edge states and a $\mathbb Z_4$ topological invariant. This is in clear contrast to existing proposals of realizing SPT phases with bosonic edge modes~\cite{phasePRL,phaseEPL,phases3}, which are protected by the $SU(N)/Z_N$ symmetry and can be realized in pure bosonic systems. The interacting fermionic SPT phase is also fundamentally different from non-interacting fermionic SPT phases for their distinct topological invariants and classifications. We numerically work out the phase diagram and propose that the interaction-induced topological phase transitions can be probed by detecting the local occupation of the clock states at the edges. Our results open up the avenue of simulating interacting fermionic SPT phases using cold atomes, and studying their classifications.

\begin{figure}[tbh]
\centering
\includegraphics[width =8cm]{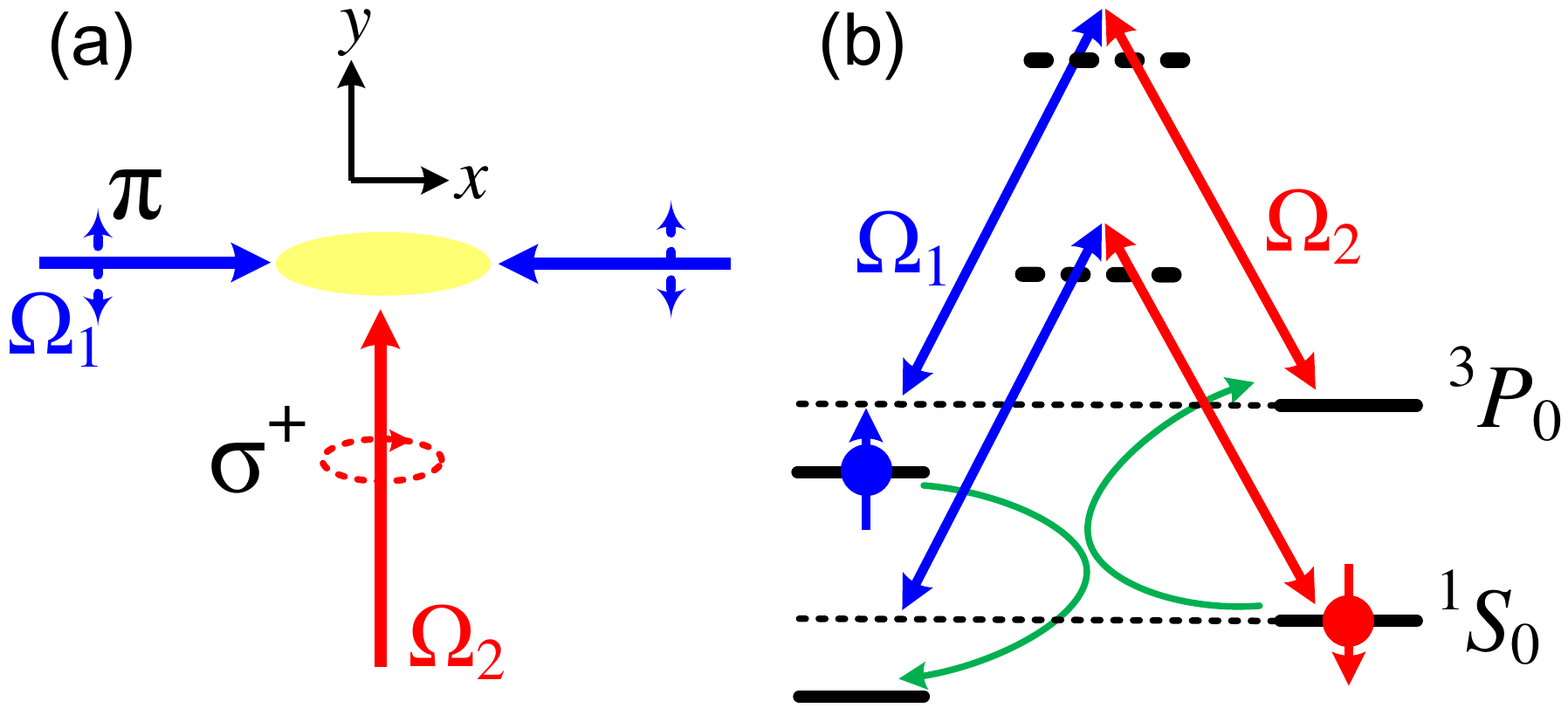}
\caption{Schematics of the system setup. (a) A quasi-1D atomic gas is subject to Raman lasers. (b) Raman level schemes in the clock-states manifold. The green curve indicates the inter-orbital spin-exchange interaction. The four nuclear spin states from $^1S_0$ and $^3P_0$ manifolds are isolated from the other nuclear spins, which can be achieved by imposing spin-dependent laser shifts~\cite{AEnew8,gyuboong}.}
\label{fig:fig1}
\end{figure}

\emph{Model}.--
We consider a quasi-1D cold atomic gas of alkaline-earth-like atoms trapped in an optical lattice potential along the axial direction, and tightly confined in the transverse directions. As illustrated in Fig.~\ref{fig:fig1}, a pair of blue-detuned Raman lasers simultaneously couple nuclear spin states
$\{|g\downarrow\rangle, |g\uparrow\rangle\}$ and $\{|e\downarrow\rangle,|e\uparrow\rangle\}$ in different orbitals. The Rabi frequencies of the two lasers forming the Raman processes are, respectively, $\Omega_1(x)=\Omega_1\cos (k_0 x)$ and $\Omega_2\exp(ik_0 y)$, which, in addition to imposing Raman-assisted spin-orbit couplings (SOCs) on the nuclear spins, give rise to 1D optical lattice potentials for the states $\{|\alpha\uparrow\rangle, |\alpha\downarrow\rangle\}$ ($\alpha=g,e$). When the Raman lasers are at a magic wavelength, the optical lattice potentials, as well as the effective Rabi frequencies of the Raman processes should be the same for the $|g\rangle$ and the $|e\rangle$ orbitals. Such a condition can be satisfied, for example, at the magic wavelength of $\sim 550$nm for $^{173}$Yb atoms~\cite{magic}. Under such a setup, the single-particle Hamiltonian can be written as
\begin{align}
\hat{H}_{0}&=\int dx \sum_{\alpha\sigma}\hat{\psi}^{\dag}_{\alpha\sigma}[-\frac{\hbar^2}{2m}\nabla^2+V\left( x\right)+\delta_{\alpha\sigma}]\hat{\psi}_{\alpha\sigma}\nonumber\\
&+ \int dx \sum_{\alpha}[M(x)\hat{\psi}^{\dag}_{\alpha\uparrow}\hat{\psi}_{\alpha\downarrow}+\mathrm{H.c.}],
\end{align}
where $\sigma=(\uparrow,\downarrow)$, $\psi_{\alpha\sigma}$ is the annihilation operator for atoms with spin $\sigma$ in the $\alpha$ orbital, and $\delta_{\alpha\sigma}$ denotes the differential Zeeman shifts under an external magnetic field~\cite{zeemanshift1,zeemanshift2}. The lattice potential $V(x)=V_0\cos^2(k_0x)$, and the Raman potential $M(x)=M_0\cos(k_0x)$, where both $V_0$ and $M_0$ are proportional to the AC polarizability of the clock states at the magic wavelength. With the latest experimental achievements on synthetic SOCs in alkaline-earth-like atoms~\cite{ye2016old,ye2016,fallani2016,gyuboong}, all the essential elements of the scheme are readily available.

While high-band effects are generally important for Raman-assisted lattice SOCs, a single-band tight-binding model is applicable when $M_0$ is not too large~\cite{highband1,highband2,highband3,highband4}. This can be satisfied by requiring $\Omega_1\gg \Omega_2$, which allows us to write down the single-particle tight-binding model
\begin{align}
\hat{H}_0 &=-t_{s} \! \sum_{<i,j>
\alpha}\!(\hat{c}_{i\alpha \uparrow }^{\dag }\hat{c}
_{j\alpha \uparrow } \! - \! \hat{c}_{i\alpha \downarrow }^{\dag }\hat{c}_{j\alpha
\downarrow }) \! + \! \sum_{i\alpha}\Gamma _{z}^{\alpha }(\hat{n}_{i\alpha \uparrow } \! - \! \hat{
n}_{i\alpha \downarrow })\nonumber\\
&+t_{\mathrm{so}}\sum_{i\alpha}(\hat{c}_{i\alpha
\uparrow }^{\dag }\hat{c}_{i+1\alpha \downarrow }-\hat{c}_{i\alpha \uparrow
}^{\dag }\hat{c}_{i-1\alpha \downarrow }+\mathrm{H.c.}),\label{eqn:H0}
\end{align}
where $\hat{c}_{i\alpha\sigma}$ is the annihilation operator for atoms on site-$i$ with spin $\sigma$ in the $\alpha$ orbital, $t_{s}=|\int dx\phi^{(i)}\bigr[-\frac{\hbar ^{2}}{2m}\nabla^{2}+V(x)\bigr]\phi^{(i+1)}|$, $t_{\mathrm{so}}=|\int dx\phi^{(i)}M(x)\phi^{(i+1)}|$, and $\Gamma _{z}^{\alpha }=\hbar (\delta _{\alpha \uparrow }-\delta _{\alpha \downarrow })/2$. Here $\phi^{(i)}$ is the lowest-band Wannier function on the $i$th site of the lattice potential $V(x)$.

In alkaline-earth-like atoms, as the electronic ($|g\rangle,|e\rangle$) and the nuclear-spin ($\left|\uparrow\right\rangle,\left|\downarrow\right\rangle$) degrees of freedom are decoupled in the clock-states manifold ($^1S_0$,$^3P_0$), the inter-orbital interaction at short ranges can only occur either in the electronic spin-singlet and nuclear spin-triplet channel $|-\rangle \equiv \frac{1}{2}(|ge\rangle-|eg\rangle)\otimes(\left|\downarrow\uparrow\right\rangle+\left|\uparrow\downarrow\right\rangle)$, or in the electronic spin-triplet and nuclear spin-singlet channel $|+ \rangle \equiv \frac{1}{2}(|ge\rangle+|eg\rangle)\otimes(\left|\downarrow\uparrow\right\rangle-\left|\uparrow\downarrow\right\rangle)$.
In a quasi-1D trapping potential and under a finite external magnetic field, these different scattering channels are coupled, and the interaction under the tight-binding approximation can be written as~\cite{ren2}
\begin{align}
&\hat{H}_{\rm int}=V_{\rm ex}\sum_i(\hat{c}^{\dag}_{ig\uparrow}\hat{c}^{\dag}_{ie\downarrow}\hat{c}_{ie\uparrow}\hat{c}_{ig\downarrow}+\mathrm{H.c.})\nonumber\\
&+U\sum_i(\hat{n}_{ig\uparrow}\hat{n}_{ie\downarrow}+\hat{n}_{ig\downarrow}\hat{n}_{ie\uparrow})+U_{0}\sum_{i \sigma}\hat{n}_{ig\sigma }\hat{n}_{ie\sigma },
\label{eqn:Hint}
\end{align}
where $\hat{n}_{i\alpha\sigma}=\hat{c}^{\dag}_{i\alpha\sigma}\hat{c}_{i\alpha\sigma}$, $U$ and $U_0$ are the on-site interaction strengths, and $V_{\rm ex}$ is the on-site inter-orbital spin-exchange interaction. All the on-site interaction parameters $\{V_{\rm ex},U,U_0\}$ can be tuned via the external magnetic field through the orbital Feshbach resonance, or via the transverse trapping frequencies through the confinement induced resonance~\cite{cirfr,ren2}. Note that $U_0=V_{\rm ex}+U$ at zero external magnetic field~\cite{ren2}. Importantly, the Hamiltonians in Eqs.~(\ref{eqn:H0}) and (\ref{eqn:Hint}) respect the aforementioned $U(1)\times Z_2^T$ symmetry~\cite{supp}.

\begin{figure}[tbh]
\centering
\includegraphics[width = 8.5cm]{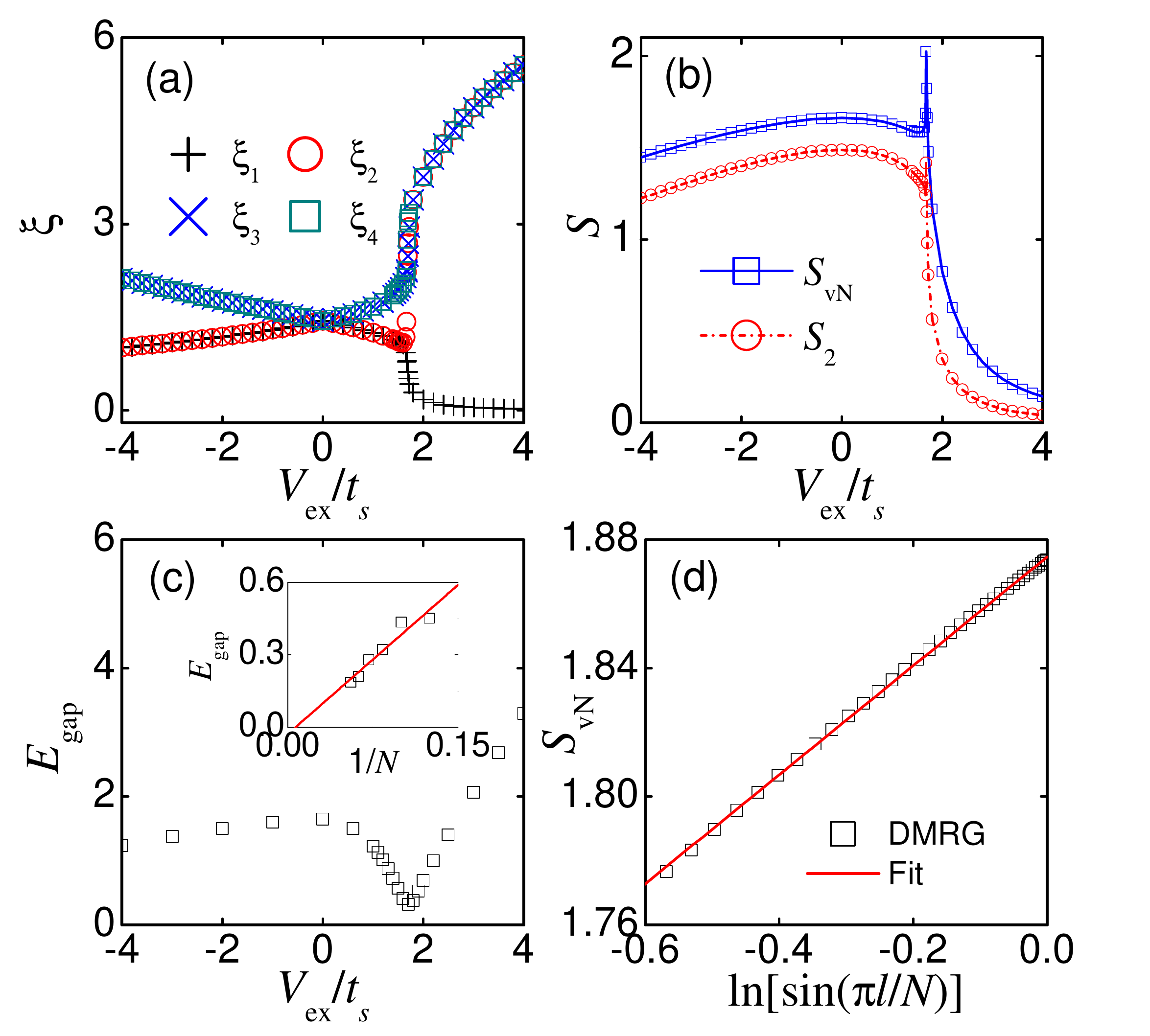}
\caption{(a) The lowest four levels in the entanglement spectrum $\protect\xi_i$ ($i=1,2,3,4$), and (b) the second-order R\'{e}nyi entropy $S_2$ and the von Neumann entropy $S_{\mathrm{vN}}$ as functions of $V_{\rm ex}/t_s$ for a chain with $N=60$ lattice sites and under open boundary conditions. (c) The bulk energy gap $E_{\rm gap}$ (the energy difference between the ground state and the first excited state) for a chain with $N=12$ lattice sites under the periodic boundary condition. (Inset) The bulk gap at the critical point as a function of $1/N$. The red solid line is a linear fit, with $E_{\rm gap}/t_s\sim -0.02\pm 0.05$ in the large-$N$ limit.  (d) The von Neumann entropy of a subchain of length $l$ as a function of $\sin(\pi l/N)$ for a chain with $N=120$ sites at the critical point $V_{\rm ex}/t_s=1.694$. The solid line is the linear fit with: $S_{\rm vN}={C\over6}\ln[\sin(\pi l/N)]+1.87$ with $C=1.018$. The central charge is six times the slope of the linear fit. All calculations are performed at half filling, and we fix $\Gamma_{z}^{g/e}=0$, $U=0$, $t_{\rm so}/t_s=0.4$.}
\label{fig:entanglement}
\end{figure}

In the absence of interactions, the ground state of the system with $\Gamma_z^{\alpha}<2t_s$ can be described by a pair of independent chiral topological insulators belonging to the AIII class~\cite{liu2013}. The topological invariant in this state is $2\in\mathbb Z$, which reflects the number of non-interacting fermionic zero modes at each edge. When symmetric perturbations, such as the interactions in Eq.~(\ref{eqn:Hint}), are turned on, the edge states are no longer non-interacting fermions but rather collective modes. The resulting SPT phase then has a $\mathbb Z_4$ invariant. This so-called $\mathbb Z_4$ reduction of the 1D class AIII systems has been discussed previously~\cite{tangwen,MFM}. We may label the $\mathbb Z_4$ phases as $K=0,1,2,3$ respectively, with $K=1$ being the root phase, i.e., a single chain of class AIII topological insulator under interactions, and $K=0$ being the trivial phase. Then the interacting SPT phase here belongs to the $K=2$ phase~\cite{supp}.
With stronger interactions, the system may be driven into trivial phases or ordered phases through continuous phase transitions, which are accompanied by the disappearance of edge modes. To understand the nature of these phases and phase transitions, we perform density matrix renormalization group (DMRG)~\cite{dmrg1,dmrg2} calculations, for which we retain $300$ truncated states per DMRG block and perform $20$ sweeps with a maximum truncation error $\sim 10^{-7}$.

\emph{Interaction-driven topological phase transition}.--
A common practice to identify 1D non-trivial topological phases is by examining the degeneracy in the entanglement spectrum of the ground state, which is defined as $\xi _{i}=-\ln (\rho _{i})$~\cite{Zhao2015,Yoshida2014,Turner2011,Pollmann2010, Fidkowski2010,Flammia2009,Li2008}. Here $\rho _{i}$ is the eigenvalue of the reduced density matrix $\hat{\rho} _{L}=\mathrm{Tr_{R}|\psi \rangle \langle \psi |}$, where $|\psi\rangle$ is the ground state and $L,R$ correspond to the left or the right half of the 1D chain. As the entanglement spectrum $\xi _{i}$ resembles the energy spectrum of edge excitations, the system is topologically non-trivial if and only if each eigenvalue $\xi_i$ is degenerate~\cite{Li2008}. We first study the case of increasing the spin-exchange interaction $V_{\rm ex}$ while fixing $U$, $\Gamma_z^{\alpha}$, and $t_{\rm so}$. In Fig.~\ref{fig:entanglement}(a), we show the four lowest levels in the entanglement spectrum as functions of $V_{\rm ex}/t_{s}$. While there is a four-fold degeneracy for the eigenstates in the entanglement spectrum with $V_{\rm ex}=0$, the degeneracy is partially lifted in the presence of weak $V_{\rm ex}$. As the degeneracy of the entanglement spectrum is generally equal to the dimension of the irreducible projective representation of the symmetry group, the lift of degeneracy can be understood as the reduction of the projective representations into irreducible ones~\cite{supp}.
For repulsive interactions ($V_{\rm ex}>0$), the entanglement spectrum is no longer degenerate beyond a critical interaction strength $V^c_{\rm ex}/t_s\sim 1.69$. Since no local-symmetry-breaking order is found on either side of the critical point, it signals a topological phase transition from an interacting SPT phase to a trivial one. For attractive interactions ($V_{\rm ex}<0$), the non-trivial SPT state persists even at large $|V_{\rm ex}|$.

The existence and the location of the interaction-driven topological phase transition can be further confirmed by entropy and bulk-gap calculations. As demonstrated in Fig.~\ref{fig:entanglement}(b), sharp features emerge at the critical point in both the second-order R\'{e}nyi entropy
$S_{2}=-\log \mathrm{Tr{(\hat{\rho} _{L}^{2})}}$~\cite{Flammia2009,Li2008,Hastings2010,Daley2012,Abanin2012,jiang2012,Islam2015},
and the von Neumann entropy $S_{\mathrm{vN}}=-\mathrm{Tr_{L}[\hat{\rho} _{L}\log \hat{\rho} _{L}]}$. In Fig.~\ref{fig:entanglement}(c), we show the bulk gap of a finite lattice with $N=12$ under periodic boundary conditions at half filling. As the system goes across the critical point, the bulk gap closes in the thermodynamic limit (inset) and opens up again, which is typical of a continuous topological phase transition.

The divergence of the von Neumann entropy at the critical point not only indicates a continuous transition but also yields the central charge, which reflects the universality class of the phase transition. In Fig.~\ref{fig:entanglement}(d), the von Neumann entropy of a subchain of length $l$ is plotted as a function of $\ln[\sin(\pi l/N)]$. The slope at large distance gives the central charge $C$~\cite{centralcharge1,centralcharge2}. From Fig.~\ref{fig:entanglement}(d), we estimate $C\sim 1.018$, which is close to the universality class of Luttinger liquid with $C=1$. Furthermore, the spin-spin correlation $\langle \hat{S}_{i\alpha x}\hat{S}_{j\alpha x}\rangle$ ($\hat{S}_{i\alpha x}$ is the $\alpha$-orbital spin operator along $x$ on site $i$) exhibits a power-law decay at the critical point, with a coefficient $\sim 1.38$~\cite{supp}. A similar power-law decay, with a coefficient $\sim 2.1$, exists for correlations of the on-site density difference between the two orbitals, which can be regarded as the spin-spin correlation in the orbital degrees of freedom~\cite{supp}. These results suggest that the system is a Luttinger liquid at the critical point.

\begin{figure}[tbh]
\centering
\includegraphics[width = 8.5cm]{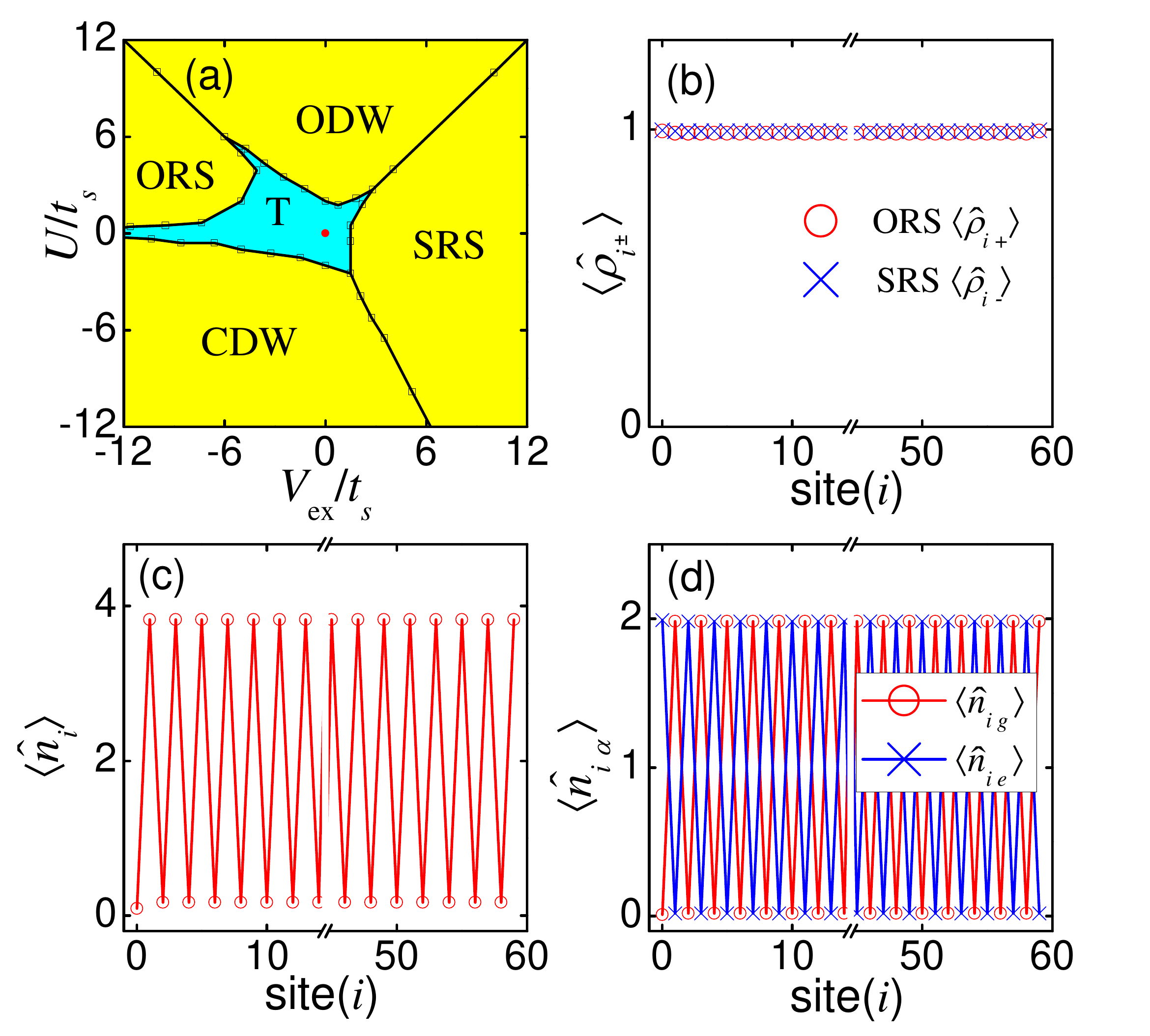}
\caption{(a) Phase diagram for a lattice with $N=60$ sites at half filling, with $\Gamma^{\alpha}_z=0$. (b) On-site densities of the states $|\pm\rangle$, with $V_{\rm ex}/t_s=-11$, $U/t_s=5$ for ORS, and $V_{\rm ex}/t_s=8$, $U=0$ for SRS, respectively. (c) Local atomic densities in the CDW state, with $\hat{n}_i=\sum_{\alpha\sigma}\hat{n}_{i\alpha\sigma}$, $V_{\rm ex}=0$, $U/t_s=-4$. (d) Local densities for different orbitals in the ODW state, with $\hat{n}_{i\alpha}=\sum_{\sigma}\hat{n}_{i\alpha\sigma}$, $V_{\rm ex}=0$, $U/t_s=8$, $t_{\rm so}/t_s=0.4$. }
\label{fig:phasediag}
\end{figure}

\emph{Phase diagram and the trivial states}.--
With the help of entanglement-spectrum and entropy calculations, we map out the phase diagram in Fig.~\ref{fig:phasediag}(a). We further identify different trivial states such as the rung-singlet states, the charge-density wave (CDW) state, and the orbital-density wave (ODW) state by calculating their corresponding local quantities~\cite{phases1,phases2,phases3}. As we have discussed previously, when $U=0$, the many-body ground state of the system can undergo a topological phase transition from a topological (T) phase to a trivial symmetric state. We define the singlet states in the orbital- (spin-) degrees of freedom as $|\pm\rangle=(|g\uparrow;e\downarrow\rangle\pm |g\downarrow;e\uparrow\rangle)/\sqrt{2}$, and analyze the local quantity $\langle \hat{\rho}_{i\pm}\rangle$, where $\hat{\rho}_{i\pm}=|\pm\rangle\langle\pm|$. As indicated in Fig.~\ref{fig:phasediag}(b), the trivial symmetric state for the repulsive $V_{\rm ex}$ case (with $U=0$) is a spin rung-singlet (SRS) state~\cite{phases3}, which can be described by the direct-product state $\prod_i |-\rangle_i$. As $U$ becomes finite, the system can become the orbital rung-singlet (ORS) state ($\prod_i |+\rangle_i$), the CDW, or an ODW state beyond the corresponding topological phase boundaries. Both the CDW and ODW are ordered trivial states with spontaneously broken chiral symmetry, which can be confirmed by calculating the corresponding local quantities as shown in Fig.~\ref{fig:phasediag}(c)(d). In Fig.~\ref{fig:phasediag}(a), we have fixed $t_{\rm so}/t_s=0.4$. If we start from the T state and decrease $t_{\rm so}$, a topological phase transition will occur at a critical $t^c_{\rm so}$ such that the system is topologically trivial for $t_{\rm so}<t^c_{\rm so}$~\cite{supp}. This highlights the role of SOC in stabilizing the T state. We also note that similar phase diagrams can be obtained at finite Zeeman fields $\Gamma^{\alpha}_z$.

\begin{figure}[tbp]
\centering
\includegraphics[width = 9cm]{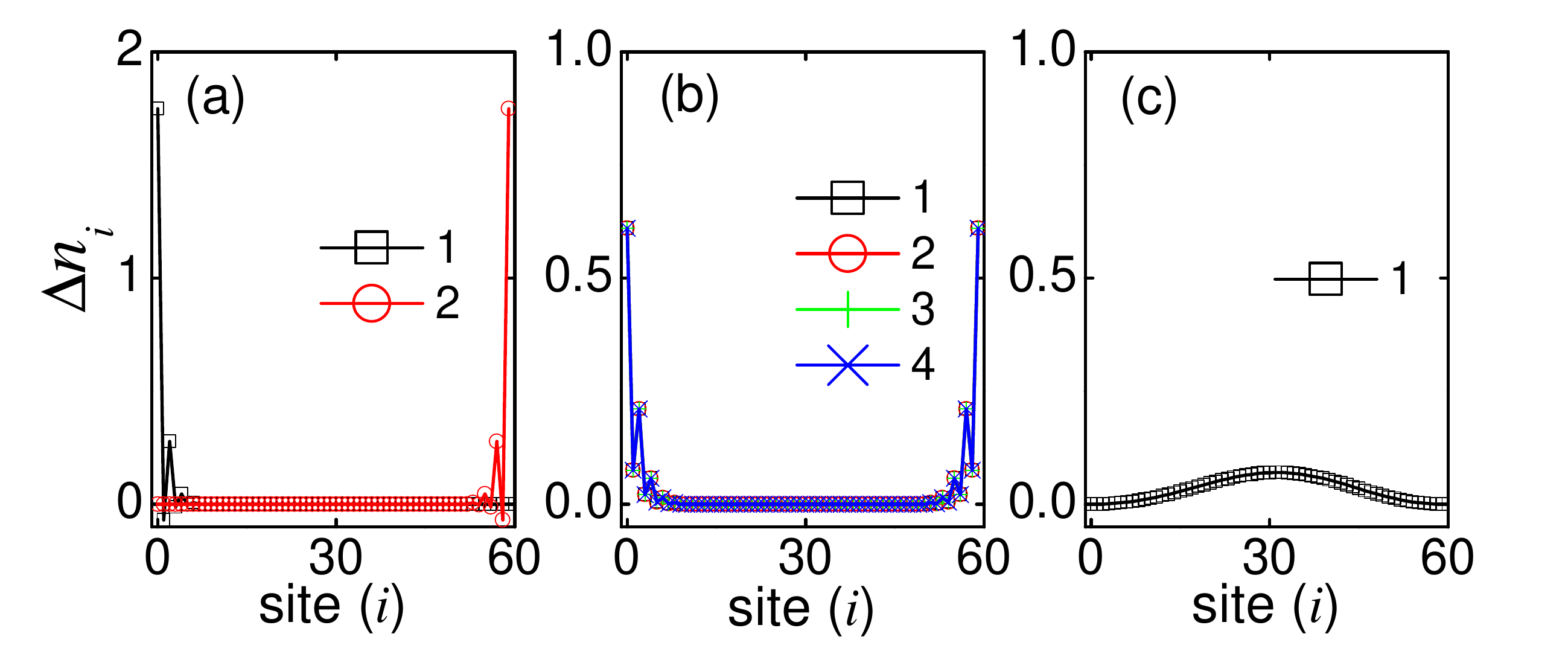}
\caption{ The edge-mode density distributions $\Delta n_i$ for (a) the T state with $V_{\rm ex}/t_s=-1$, (b) the T state with $V_{\rm ex}/t_s=1$, and (c) the SRS state with $V_{\rm ex}/t_s=2$. Other parameters are the same as those in Fig.~\ref{fig:entanglement}. The numbers in the figure legends label different degenerate ground states.}
\label{fig:edge}
\end{figure}

\emph{Topological edge modes}.--
Another prominent feature of SPT states is the existence of topological edge modes at the boundaries, which are robust against symmetric perturbations. Whereas the existence (absence) of the topological edge modes signals that the system is topologically non-trivial (trivial), the topological edge modes in the interacting SPT state significantly affect the degeneracy of the many-body ground state at half filling.
In the absence of interactions, the ground state should be six-fold degenerate, when half of the four spin-polarized fermionic edge modes (two for each edge) are occupied.
Under finite interactions, we numerically confirm that the ground-state degeneracy is dependent on the bulk interactions, which is governed by the projective representations of the symmetry group~\cite{supp}.
For repulsive bulk interactions ($V_{\rm ex}>0$), the ground state is four-fold degenerate, with only one edge mode occupied on each edge. For attractive bulk interactions ($V_{\rm ex}<0$), the ground state is two-fold degenerate, with the two zero modes at each edge either both empty or both occupied.

\emph{Detection}.--
The understanding of the topological edge modes not only provides a way to manipulate the edge modes by tuning the bulk interactions, but also offers a detection scheme based on measuring the localized density of edge modes~\cite{ladderspt}.
The density distribution of the edge modes can be calculated as $\Delta n_i=\sum_{\alpha \sigma}\langle \hat{n}_{i\alpha\sigma}(2N)-\hat{n}_{i\alpha\sigma}(2N-2)\rangle$, where $\langle \hat{n}_{i\alpha\sigma}(M)\rangle$ is the expectation value of the density operator $\hat{n}_{i\alpha\sigma}$ in the ground state for $M$ atoms on $N$ lattice sites. As shown in Fig.~\ref{fig:edge}, while the trivial SRS state features a non-local density distribution with negligible population at the edges, the T states with two- and four-fold ground-state degeneracies feature localized density distributions at the edges. In particular, for the T states with two-fold ground-state degeneracy, the localized edge-mode density is twice as that with four-fold degeneracy. We have checked that other trivial states have similarly negligible population at the edges as the SRS. Thus, the localized density of the edge modes allows for the detection of the interaction-driven topological phase transition, and enables the determination of the edge-mode degeneracy in the T state. To measure the local atomic density at the edges, one needs to measure the overall occupation of the $|g\rangle$ and the $|e\rangle$ orbitals at the boundaries, which can be achieved, respectively, by coupling the $^1S_0$-$^1P_1$ and the $^3P_0$-$^1P_1$ (or the $^3P_0$-$^3S_1$-$^1P_1$ two-photon) transitions and recording the resulting fluorescence. To selectively apply local operations, a localized laser field can be applied at the edges to provide the necessary energy shifts. Finally, we note that the topological phase transition may also be detected by probing the non-local string-order parameters~\cite{quella12}, which can be achieved, for instance, by measuring single-site-resolved on-site parity of the atom number~\cite{blochstringorder}.

\emph{Final remarks}.--
We have proposed to use alkaline-earth-like atoms to investigate SPT states for interacting fermions, which are induced by SOC and the inter-orbital spin-exchange interactions in these atoms. An alternative scheme with separate Raman lasers for different orbitals can also be considered, where the reflection symmetry between the two orbitals would be broken even at a zero magnetic field. However, the non-trivial SPT should survive, as it is not protected by the inter-orbital reflection symmetry.

\emph{Acknowledgments}.--
We thank Gyu-Boong Jo and Thomas Quella for helpful comments. This work is supported by the National Key R\&D Program (Grant Nos. 2017YFA0304203 and 2016YFA0301700), the NKBRP (2013CB922000), the National Natural Science Foundation of China (Grant Nos. 60921091, 11274009, 11374283, 11422433, 11434007, 11434011, 11522436, 11522545, 11574392, 11674200), and the Research Funds of Renmin University of China (10XNL016, 16XNLQ03). X. Z. and G. C. are supported by the PCSIRT under Grant No. IRT13076, the FANEDD under Grant No. 201316, SFSSSP; OYTPSP, and SSCC. W. Y. acknowledges support from the ``Strategic Priority Research Program(B)'' of the Chinese Academy of Sciences, Grant No. XDB01030200.

\emph{Note added}.--
After the submission of our work, an experiment appeared online (arXiv:1706.00768) realizing non-interacting class AIII SPT state in $^{173}$Yb atoms. The experiment paves the way toward implementing the interacting SPT discussed here.

\newpage
\begin{widetext}
\appendix

\renewcommand{\thesection}{\Alph{section}}
\renewcommand{\thefigure}{S\arabic{figure}}
\renewcommand{\thetable}{S\Roman{table}}
\setcounter{figure}{0}
\renewcommand{\theequation}{S\arabic{equation}}
\setcounter{equation}{0}

\section{Supplemental Materials}

In this Supplemental Materials, we provide details on the topological edge
modes, the characterization of the topological phase transitions, the
topological invariant of the interacting fermionic SPT state, and the effect
of spin-orbit coupling. The notation used here follows that of the main text.

\subsection{\textbf{A. Topological edge modes and the chemical potential}}

\label{supp:edge} We calculate the chemical potential $%
\mu(M)=E_{0}(M+1)-E_{0}(M)$, which is essentially the energy required to add
an atom to a system of $M$ atoms. Here, $E_0(M)$ is the ground-state energy
of $M$ atoms on $N$ lattice sites with open boundary conditions. As
illustrated in Fig.~\ref{fig:supp1}(a), when the system is in the SPT state,
mid-gap modes associated with the topological edge modes emerge in the
chemical-potential spectrum. These mid-gap modes represent the excitation
energies required as the occupation number of the four topological edge
modes (two for each orbital) increases sequentially from zero to four. As $%
V_{\mathrm{ex}}$ increases, these mid-gap modes would shift toward the bulk
and eventually merge into the bulk spectrum. Such a behavior can be
characterized by calculating the excitation energy gap between the bulk
spectrum and the nearest mid-gap state: $\Delta%
\mu=[E_0(2N+3)-E_0(2N+2)]-[E_0(2N+1)-E_0(2N)]$, where $2N$ corresponds to
the half-filling condition for a lattice with $N$ sites. As illustrated in
Fig.~\ref{fig:supp1}(b), the excitation gap vanishes at the critical point,
indicating the merging of the mid-gap modes into the bulk spectrum.

\begin{figure}[tbh]
\centering
\includegraphics[width = 10cm]{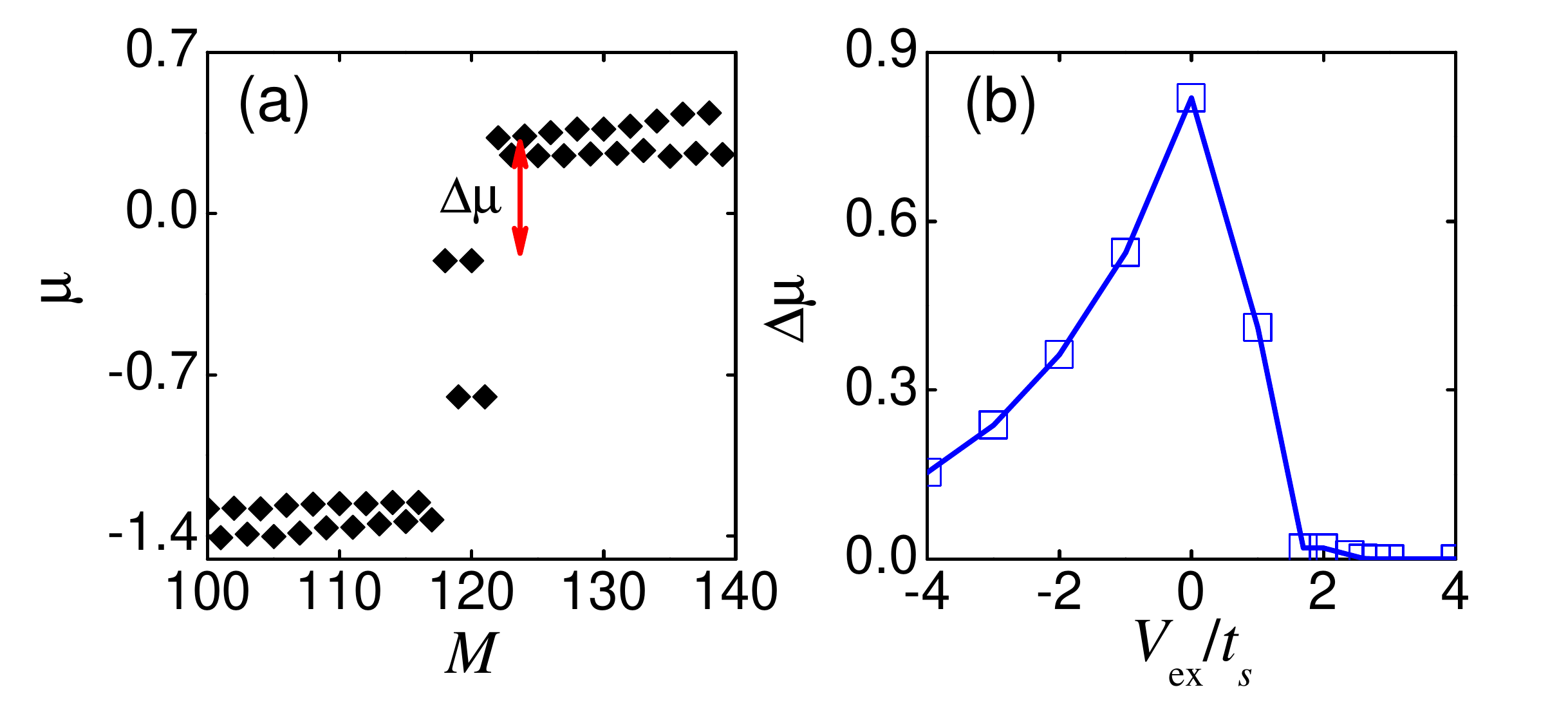}
\caption{(a) Chemical potential $\protect\mu(M)=E_{0}(M+1)-E_{0}(M)$ as
functions of $V_{\mathrm{ex}}/t_s$ for a chain with $N=60$ lattice sites at
half filling with the atom number $M=120$, and under open boundary
conditions. Here, $V_{\mathrm{ex}}/t_s=-1$. (b) The excitation energy gap $%
\Delta \protect\mu$ as a function of $V_{\mathrm{ex}}/t_s$. For both plots,
we have $\Gamma_{z}^{g/e}=0$, $U=0$, $U_0/t_s=-1$.}
\label{fig:supp1}
\end{figure}

\subsection{\textbf{B. Characterizing the topological phase transitions}}

\label{supp:corr} The continuous topological phase transition between the T
state and the rung-singlet states can be characterized by the relevant
correlation functions at the critical point. Taking the T-SRS boundary as an
example, we find that the correlation of the on-site density difference
between the two orbital, defined as $\langle \hat{n}_{g-e}\hat{n}%
_{g-e}\rangle =\langle (\hat{n}_{ig}-\hat{n}_{ie})(\hat{n}_{i+d,g}-\hat{n}%
_{i+d,e})\rangle $, exhibits a power-law decay at the phase boundary.
Similarly, the spin-spin correlation, $\langle \hat{S}_{\alpha }^{x}\hat{S}%
_{\alpha }^{x}\rangle =\langle \hat{S}_{i,\alpha x}\hat{S}_{i+d,\alpha
x}\rangle $, also decays in a power-law fashion at the critical pint. Here, $%
\hat{S}_{i,\alpha x}=\sqrt{2}/2(\hat{c}_{i\alpha \uparrow }^{\dag }\hat{c}%
_{i\alpha \downarrow }+$H.c.$)$. In Fig.~\ref{fig:supp2}, we show the linear
fit on a log-log plot of the correlation functions versus the distance
between sites $d$. The power-law decay of the correlation functions,
combined with the central charge calculation, suggest that the system is a
Luttinger liquid at the critical point.

\begin{figure}[tbp]
\centering
\includegraphics[width = 10cm]{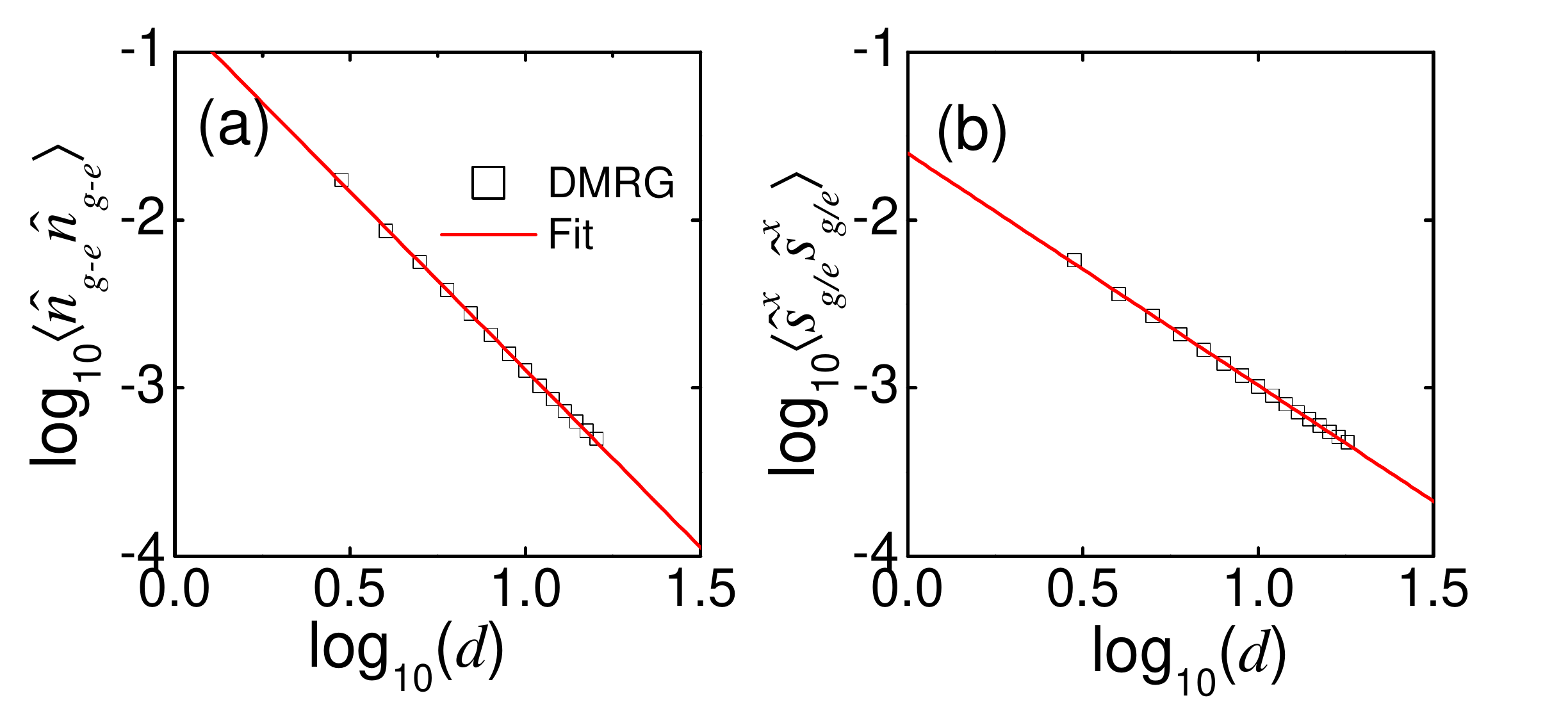}
\caption{(a) Power-law decay of the correlation of the density-difference at
the critical point in Fig.~2(d) of the main text. The linear fit is given by
$y=-2.1x-0.77$. (b) Power-law decay of the spin-spin correlation at the same
critical point. The linear fit is given by $y=-1.38x-1.6$. }
\label{fig:supp2}
\end{figure}

Similarly, we find that the phase transitions between the T state and the
CDW, the ODW, and the ORS states are all continuous. As shown in Fig.~\ref%
{fig:supp3}, by characterizing the divergence of the von Neumann entropy at
the corresponding phase boundaries, we find that the central charges for the
T-ORS, the T-ODW, and the T-CDW phase transitions are $\sim 0.978$, $\sim
0.972$, and $\sim 1.008$, respectively.

\begin{figure}[tbp]
\centering
\includegraphics[width = 14cm]{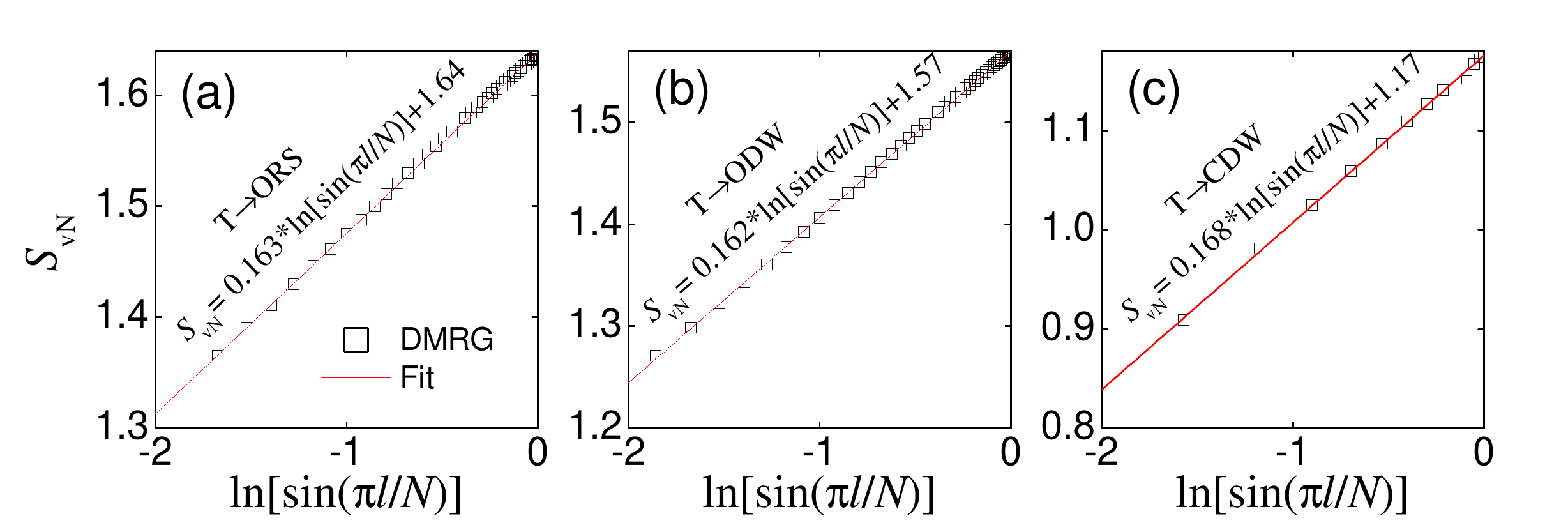}
\caption{The von Neumann entropy of a subchain of length $l$ as a function
of $\sin(\protect\pi l/N)$ for a chain with $N=100$ sites. (a) At the T-ORS
phase boundary with $V_{\mathrm{ex}}/t_s=-4.982$, $U/t_s=2$; (b) at the
T-ODW phase boundary with $V_{\mathrm{ex}}/t_s=-1.190$, $U/t_s=2.791$; (c)
at the T-CDW boundary with $V_{\mathrm{ex}}/t_s=-4.995$, $U/t_s=-0.995$.}
\label{fig:supp3}
\end{figure}

\subsection{\textbf{C. Topological invariant of the SPT state}}

\label{supp:proj}

In this section, we discuss the topological invariant of the interacting
fermionic SPT phase protected by the $U(1)\times Z_2^T$ symmetry as
discussed in the main text. Generally, the topological invariants and
classifications of short-range entangled interacting fermionic SPT phases
with particle number conservation is not well established in two and higher
dimensions. However, in one dimension, since fermionic systems can be mapped
into bosonic systems via the Jordan-Wigner transformation, the topological
invariants of bosonic SPT phases, while different from those of the
corresponding fermionic SPT phases, can be applied for fermionic systems
with slight modifications~\cite{chenguwen}.

For one-dimensional bosonic SPT phases, the topological invariants are the
equivalence classes of projective representations of the symmetry group
carried by the edge states. For the $U(1)\times Z_2^T$ symmetry, the
invariants of the SPT phases form a $\mathbb{Z}_2\times \mathbb{Z}_2$ group
according to the second group cohomology $\mathcal{H}^2(U(1)\times Z_2^T,
U(1))=\mathbb{Z}_2\times \mathbb{Z}_2$. In other words, the classification
of the corresponding bosonic SPT phases is $\mathbb{Z}_2\times \mathbb{Z}_2$%
. In the following, we will calculate the projective representations carried
by the edge states of femrionic SPT phases with the same symmetry group. In
contrast to the bosonic SPT phases, the fermionic SPT phases have a $\mathbb{%
Z}_4$ topological invariant owning to the fermionic exchanging statistics of
the femionic edge zero modes.

As discussed in the main text, for a single chain, i.e. with only one
orbital degree of freedom, the left edge mode $|\psi_l\rangle$ is an
eigenstate of the spin operator $\hat{S}_x=\hat{\sigma}_x/2$ with the
eigenvalue $+1/2$ (we have taken $\hbar$ to be one), and the right edge mode
$|\psi_r\rangle$ is an eigenstate of $\hat{S}_x$ with the eigenvalue $-1/2$,
\begin{eqnarray}
\hat{S}_x|\psi_l\rangle = {\frac{1}{2}}|\psi_l\rangle,\ \ \hat{S}%
_x|\psi_r\rangle = -{\frac{1}{2}}|\psi_r\rangle.
\end{eqnarray}

So we can describe the edge modes by fermion operators $\hat{c}_{1l}$ and $%
\hat{c}_{1r}$, $|\psi_{l}\rangle = \hat{c}_{1l}^\dag |\mathrm{vac}\rangle$, $%
|\psi_{r}\rangle = \hat{c}_{1r}^\dag |\mathrm{vac}\rangle$ respectively.
Here the subscript $1$ indicates the orbital degree of freedom.

Without loss of generality, we only consider the left boundary. The edge
mode varies under symmetry action as the following:
\begin{eqnarray}
&&\hat U(\theta)\hat{c}_{1l}\hat U(\theta)^{-1} = e^{i\theta} \hat{c}_{1l},
\label{SymU} \\
&&\hat{(\mathcal{CT})}\hat{c}_{1l}\hat{(\mathcal{CT})}^{-1} = \hat{c}%
_{1l}^\dag.  \label{SymT}
\end{eqnarray}

The fermion mode $\hat{c}_{1l}$ spans a two-dimensional Hilbert space $%
|0\rangle$ and $|1\rangle = \hat{c}_{1l}^\dag |0\rangle$. In this
two-dimensional Hilbert space of the left edge states, we can replace $\hat{c%
}_{1l}$ and $\hat{c}_{1l}^\dag$ as $\hat{\sigma}^- = (\hat{\sigma}_x-i\hat{%
\sigma}_y)/2$ and $\hat{\sigma}^+=(\hat{\sigma}_x+i\hat{\sigma}_y)/2$,
respectively. From Eqs. (\ref{SymU}) and (\ref{SymT}), we can solve the
matrix form of symmetry action as
\begin{eqnarray}
&&\hat U(\theta)\to M_1(\theta)=\left(
\begin{matrix}
1 & 0 \\
0 & e^{i\theta}%
\end{matrix}%
\right), \\
&&\hat{(\mathcal{CT})}\to M_1\hat{(\mathcal{CT})} = \hat{\sigma}_x K.
\end{eqnarray}

Since $M_1\hat{(\mathcal{CT})}K M_1(\theta)= e^{-i\theta}M_1(\theta) M_1\hat{%
(\mathcal{CT})}K$, these matrices form a projective representation of the
symmetry group $U(1)\times Z_2^T$, and they describe how the edge state vary
under the action of the symmetry operations.

Now we consider a two-leg ladder, i.e. with both orbital degrees of freedom,
as discussed in the main text. The edge modes of the second chain located at
the left and the right edge can be described as $\hat{c}_{2l}$ and $\hat{c}%
_{2r}$, respectively. Now at the left edge we have two fermionic zero modes $%
\hat{c}_{1l}$ and $\hat{c}_{2l}$. The Hilbert space spanned by the left edge
states is four-dimensional. Since $\{\hat{c}_{1l}, \hat{c}_{2l}\}=0$,
following the concept of the Jordan-Wigner transformation the fermion
operators can be written in matrix form as
\begin{eqnarray}
\hat{c}_{1l} = I \otimes \hat{\sigma}^-,\ \ \hat{c}_{2l} = \hat{\sigma}^-
\otimes \hat{\sigma}^z.
\end{eqnarray}
Since the symmetry acts on the fermion operators as follows,
\begin{eqnarray}
&&\hat U(\theta)\hat{c}_{\alpha l} \hat U(\theta)^{-1} = e^{i\theta} \hat{c}%
_{\alpha l},  \label{SymU2} \\
&&\hat{(\mathcal{CT})}\hat{c}_{\alpha l}\hat{(\mathcal{CT})}^{-1} = \hat{c}%
_{\alpha l}^\dag,  \label{SymT2}
\end{eqnarray}
we can solve the representation matrix of the symmetry operations as
\begin{eqnarray}
&&\hat U(\theta)\to M_2(\theta)=\left(
\begin{matrix}
1 & 0 \\
0 & e^{i\theta}%
\end{matrix}%
\right)\otimes \left(
\begin{matrix}
1 & 0 \\
0 & e^{i\theta}%
\end{matrix}%
\right),  \label{4DU} \\
&&\hat{(\mathcal{CT})}\to M_2\hat{(\mathcal{CT})} = \hat{\sigma}_y\otimes
\hat{\sigma}_x K.
\end{eqnarray}
The above matrices can be simultaneously block-diagonalized as
\begin{eqnarray}
&&M_2(\theta)=\left(%
\begin{matrix}
1 &  &  &  \\
& e^{2i\theta} &  &  \\
&  & e^{i\theta} &  \\
&  &  & e^{i\theta}%
\end{matrix}%
\right), \\
&& M_2\hat{(\mathcal{CT})} = \left(
\begin{matrix}
\hat{\sigma}_y &  \\
& -\hat{\sigma}_y%
\end{matrix}%
\right) K,  \label{4DT}
\end{eqnarray}
which is a direct sum of two irreducible projective representations (of the
same class).

At half filling, only two of the four edge modes are occupied. The ground
state should then be six-fold degenerate in the absence of interactions. If
the effective interaction between the two legs is repulsive, there is always
one fermion zero mode occupied at one edge, which varies as the irreducible
projective representation under symmetry action,
\begin{eqnarray}
&&M_{2}(\theta )=\left(
\begin{matrix}
e^{i\theta } &  \\
& e^{i\theta }%
\end{matrix}%
\right) ,  \label{2DU1} \\
&&M_{2}\hat{(\mathcal{CT})}=-\hat{\sigma}_{y}K.  \label{2DT1}
\end{eqnarray}%
In this case, the ground state at half filling features a four-fold
degeneracy. On the other hand, if the interaction is attractive, the fermion
zero modes are either not occupied or doubly occupied at each edge, they
carry the following irreducible projective representation,
\begin{eqnarray}
&&M_{2}(\theta )=\left(
\begin{matrix}
1 &  \\
& e^{2i\theta }%
\end{matrix}%
\right) ,  \label{2DU2} \\
&&M_{2}\hat{(\mathcal{CT})}=\hat{\sigma}_{y}K.  \label{2DT2}
\end{eqnarray}%
The ground state is then two-fold degenerate at half filling. Note that
while the projective representation above indicates four ground states, only
two of them are at half filling. By comparing numerical results with the
theory above, we find that the repulsive effective interaction corresponds
to a repulsive bulk spin-exchange interaction $V_{\mathrm{ex}}>0$, and the
attractive effective interaction corresponds to an attractive bulk
interaction $V_{\mathrm{ex}}<0$. This implies that when the bulk
spin-exchange interactions are tuned, the ground-state degeneracy at half
filling can change from two (for $V_{\mathrm{ex}}<0$) to six (for $V_{%
\mathrm{ex}}=0$) and finally to four (for $V_{\mathrm{ex}}>0$). This is
illustrated in Fig.~\ref{fig:supp4} for a finite chain with $N=10$ at half
filling.

\begin{figure}[tbp]
\centering
\includegraphics[width = 17cm]{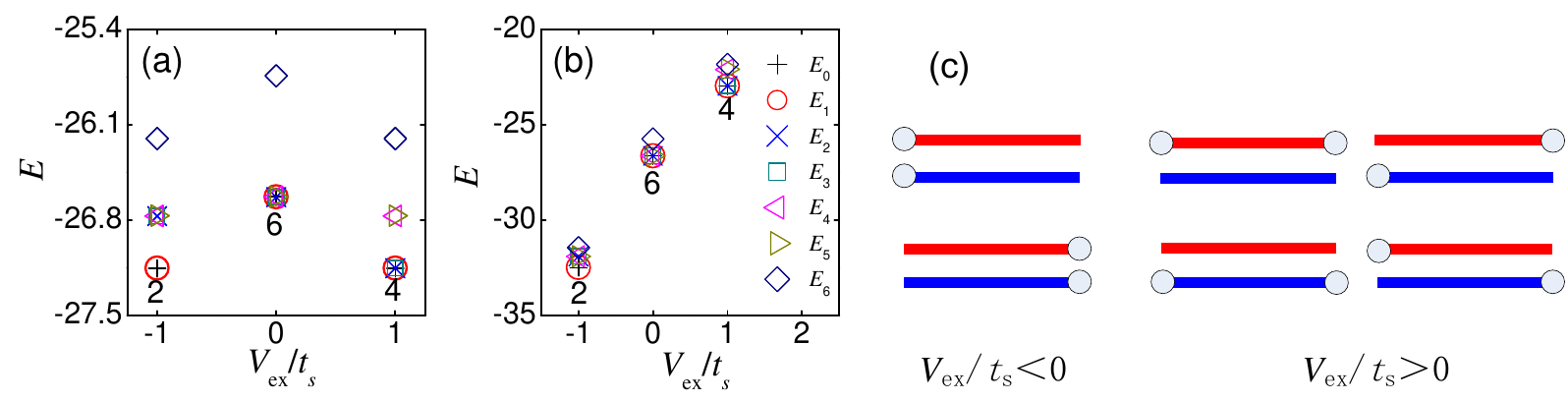}
\caption{The ground-state degeneracy for a finite chain with $N=10$ at half
filling. (a) $U_0=0$ and (b) $U_0=V_{\mathrm{ex}}+U$. We fix $U=0$ here. (c)
A schematic illustration of the occupied edge fermion modes of the
degenerate ground states for different bulk interactions.}
\label{fig:supp4}
\end{figure}

Similarly, if one more chain is stacked into the ladder, then the resultant
edge states carry the following irreducible projective representation,
\begin{eqnarray}
&&M_{3}(\theta )=\left(
\begin{matrix}
e^{im\theta } &  \\
& e^{in\theta }%
\end{matrix}%
\right) , \\
&&M_{3}\hat{(\mathcal{CT})}=\hat{\sigma}_{y}K,
\end{eqnarray}%
with $m+n=$odd. Finally, if four chains are stacked together, then the edge
states can be fully gapped out and it is described by a trivial
one-dimensional irreducible projective representation.

As such, the fermionic SPT phase discussed in the main text can be prepared
by first stacking two identical chains of class AIII topological insulators
together, and then switching on the symmetry-preserving interactions. The
resulting T phase belongs to one of the three topological non-trivial states
in the $\mathbb{Z}_4$ classification. This so-called $\mathbb{Z}_4$
reduction of the one-dimensional class AIII systems has been theoretically
discussed previously~\cite{tangwen,MFM}. We may label the $\mathbb{Z}_4$
phases as $K=0,1,2,3$ respectively, with $K=1$ being the root phase, i.e., a
single chain of class AIII topological insulators with interactions turned
on, and $K=0$ being the trival phase. Then the interacting SPT phase here
belongs to the $K=2$ phase.

Since the degeneracy in the entanglement spectrum is generally equal to the
dimension of the irreducible projective representation of the symmetry
group, we now show that the partial lift of the degeneracy in the
entanglement spectrum in Fig.~2(a) of the main text can be interpreted in
terms of reduction of the reducible projective representations of the
stacked chains into irreducible ones. For a single chain, the entanglement
spectrum $\xi_i$ (see the main text for definition) are two-fold degenerate,
namely, each of the two-fold degenerate entanglement weight $\rho_i$ is
associated with a two-dimensional irreducible projective representation
(carried by the entanglement Schmidt eigenstates, the so-called virtual
states). When stacking two chains without interaction, the Schmidt
eigen-space corresponding to the weight $\Lambda_{ij}=\rho_i^{(1)}%
\rho_j^{(2)}$ is a direct product of the fermionic eigen-spaces of $%
\rho_i^{(1)}$ and $\rho_i^{(2)}$ of the two chains respectively. This
four-dimensional Hilbert space carries the four-dimensional reducible
projective representations as shown in Eqs.~(\ref{4DU})$\sim$(\ref{4DT}),
which explains the four-fold degeneracy in the entanglement spectrum at $V_{%
\mathrm{ex}}/t_s=0$. Since the four-dimensional projective representation
can be reduced to a direct sum of a pair of two-dimensional irreducible
representations, when the symmetry-reserving interactions are switched on,
the four-fold degenerate weight $\Lambda_{ij}=\rho_i^{(1)}\rho_j^{(2)}$
reduces to a pair of two-fold degenerate weights $\Lambda_{ij}^{\prime }$
and $\Lambda_{ij}^{\prime \prime }$, which carry irreducible representations
given in Eqs.~(\ref{2DU1}), (\ref{2DT1}) and (\ref{2DU2}), (\ref{2DT2})
respectively.

\subsection{\textbf{D. The effect of spin-orbit coupling}}

In our system, a key element in generating the fermionic SPT phase is the
Raman-assisted spin-orbit coupling within the orbital degree of freedom. The
effect of spin-orbit coupling is strikingly clear when we compare the phase
diagram in Fig.~3 of the main text with that of a closely related but
different system studied in Ref.~\cite{phases1}. It is shown in Ref.~\cite%
{phases1} that, in the absence of the spin-orbit coupling term $t_{\mathrm{so%
}}$, the SPT phase (T phase) should be replaced by a conventional
spin-Peierls-like (SP) phase. Such an SP state is topologically trivial, and
can be characterized by the order parameter
\begin{align}
\mathrm{SP}&={\frac{2}{N}}\sum_{i=N/4}^{3N/4} \left[(-1)^i
\sum_{\alpha\sigma} \hat c_{i+1 \alpha \sigma}^{\dag} \hat c_{i \alpha\sigma}%
\right],  \label{eqn:suppSP}
\end{align}
where $N$ is the total number lattice sites. Note that as we wish to
characterize the order parameter in the bulk, a truncation in the range of
summation over $i$ is taken to eliminate the unwanted impact of the edge
modes (more on this point later).

\begin{figure}[tbp]
\centering
\includegraphics[width = 10cm]{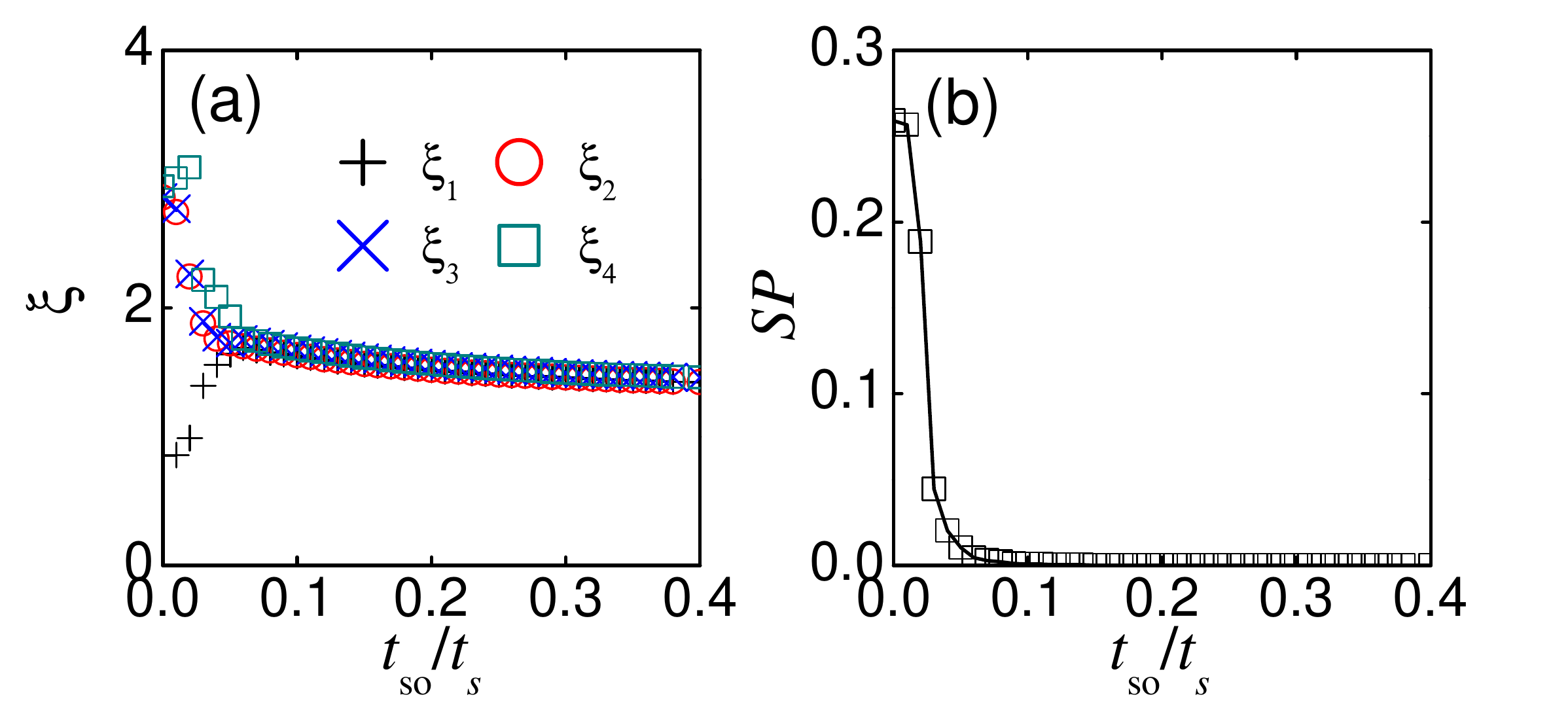}
\caption{(a) The lowest four levels in the entanglement spectrum, and (b)
the SP order parameter as functions of increasing spin-orbit-coupling
parameter $t_{\mathrm{so}}$. For both subplots we have $U/t_s=1$, $V_{%
\mathrm{ex}}/t_s=-1$, $N=64$.}
\label{fig:supp5}
\end{figure}

In Fig.~\ref{fig:supp5}, we show the lowest four levels of the entanglement
spectrum and the SP order parameter with increasing $t_{\mathrm{so}}/t_s$.
One can clearly identify a critical parameter $t^c_{\mathrm{so}}/t_s$, below
which the ground state in the entanglement spectrum is non-degenerate and
the SP order parameter is finite. Beyond $t^c_{\mathrm{so}}/t_s$, however,
the eigenstates of the entanglement spectrum become two-fold degenerate and
the SP order parameter becomes vanishingly small. This clearly indicates the
existence of a topological phase transition between the SP state at small $%
t_{\mathrm{so}}/t_s$ and the SPT state at larger $t_{\mathrm{so}}/t_s$. The
phase transition is driven by competition between the spin-orbit coupling
and the interaction.

\begin{figure}[tbp]
\centering
\includegraphics[width = 10cm]{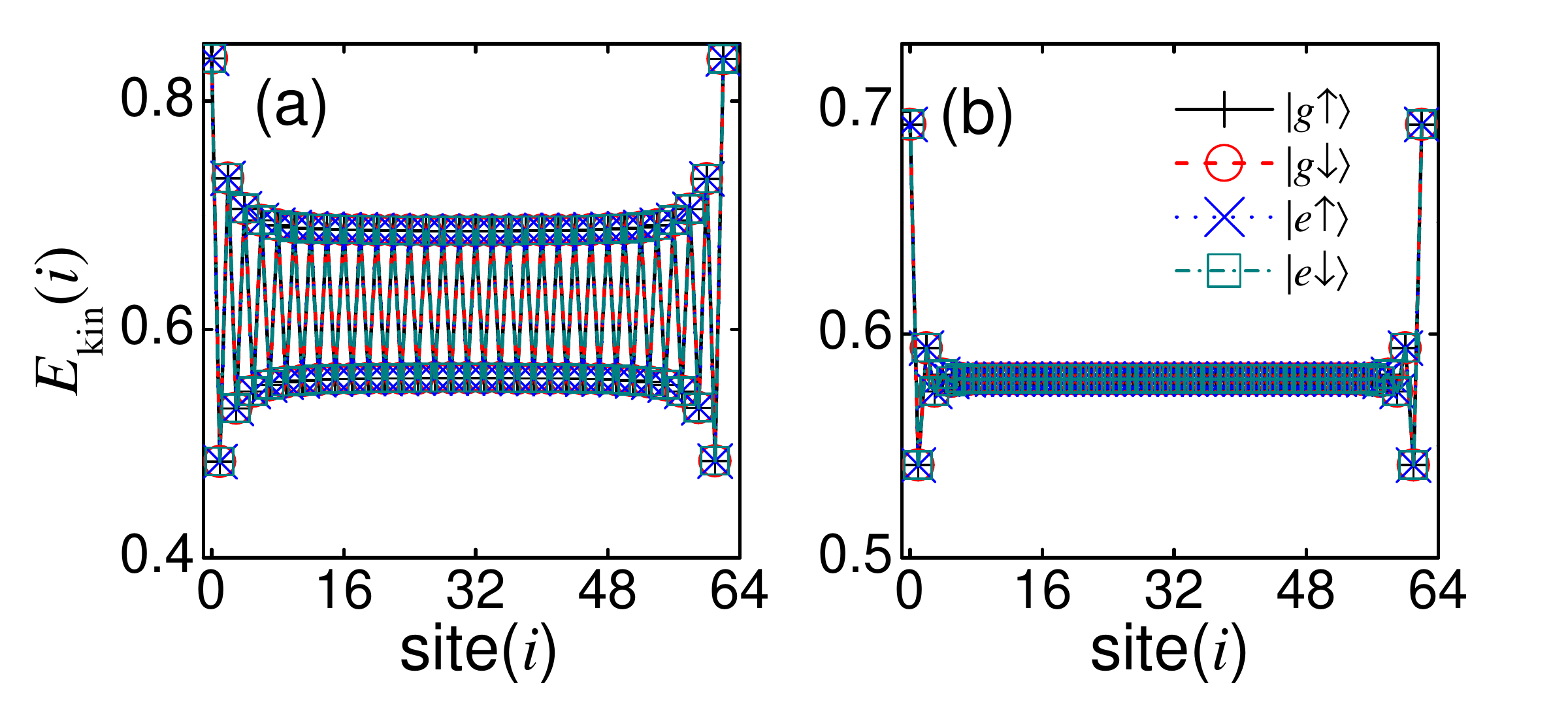}
\caption{ The kinetic energy densities for (a) the SP state with $t_{\mathrm{%
so}}=0$, and (b) the SPT state with $t_{\mathrm{so}}/t_s=0.4$. For both
subplots we have $U/t_s=1$, $V_{\mathrm{ex}}/t_s=-1$, $N=64$.}
\label{fig:supp6}
\end{figure}

We further illustrate the difference between the SPT state and the SP state
by plotting the kinetic energy density
\begin{eqnarray}
E_{\mathrm{kin}}(i)=\hat c_{i+1 \alpha\sigma}^{\dag} \hat c_{i \alpha\sigma}
\end{eqnarray}
in Fig.~\ref{fig:supp6}. While the kinetic energy density oscillates in the
bulk of the SP state, which is a signature for the dimerization typical of
the SP state, there is no such oscillation in the bulk of the SPT state.
Note that in the SPT state, the kinetic energy density only oscillates at
the two edges, which shows the impact of edge modes rather than the bulk
states. This observation is consistent with our truncation of edge sites in
the definition of the SP order parameter in Eq.~(\ref{eqn:suppSP}).

\begin{figure}[tbp]
\centering
\includegraphics[width = 9.5cm]{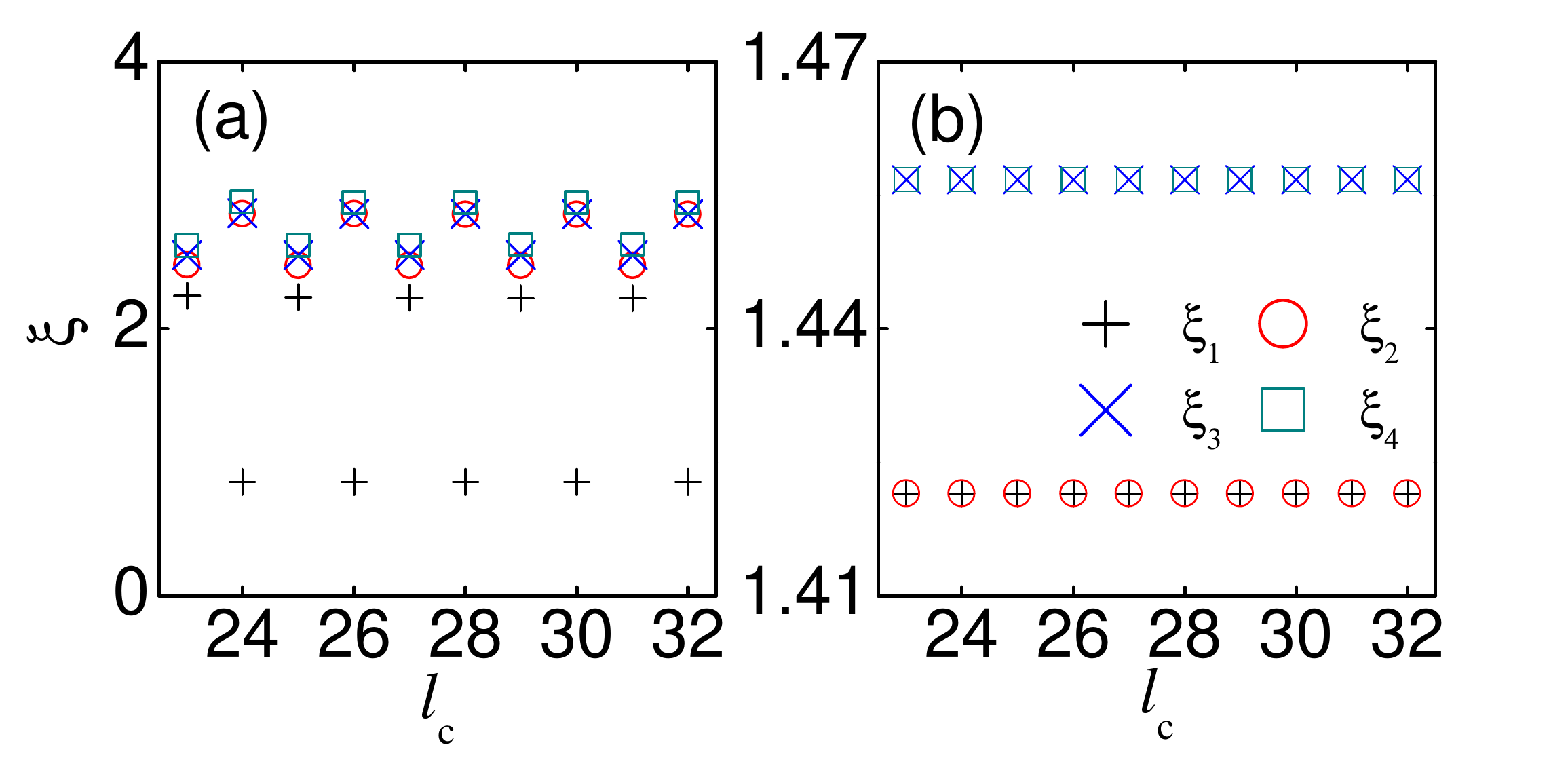}
\caption{The lowest six levels of the entanglement spectrum in (a) the SP
state with $t_{\mathrm{so}}/t_s=0$, and (b) the SPT state with $t_{\mathrm{so%
}}/t_s=0.4$ as functions of the bipartition position $l_c$. For both
subplots we have $U/t_s=1$, $V_{\mathrm{ex}}/t_s=-1$, $N=64$.}
\label{fig:supp7}
\end{figure}

The lack of dimerization in the bulk of the SPT state can also be confirmed
by varying the bipartition position $l_c$ between the left (L) and the right
(R) halves of the lattice while calculating the entanglement spectrum. In
the SP state, as illustrated in Fig.~\ref{fig:supp7}(a), the ground state of
the entanglement spectrum remains to be non-degenerate regardless of odd or
even $l_c$. In contrast, in the SPT state, as in Fig.~\ref{fig:supp7}(b),
the eigenstates in the entanglement spectrum are two-fold degenerate
regardless of the value of $l_c$. A closer observation reveals that, in the
SP state, the lowest six levels in the entanglement spectrum are very close
to one another (quasi-degenerate) when $l_c$ sits on odd sites. Such a
quasi-degeneracy originates from the dimerized nature of the SP state. As
discussed in Refs.~\cite{phases1,annphys}, the SP state is originally
defined for the $SU(4)$ Hubbard model. In such a system, the ground state of
the entanglement spectrum should be non-degenerate for even $l_c$ and
six-fold degenerate for odd $l_c$~\cite{annphys}, due to the alternating
strong and weak bonds (dimerization). In our system, the $SU(4)$ symmetry is
not fully preserved under the interaction parameters governed by the
realistic scattering parameters of $^{173}$Yb atoms. This gives rise to the
slightly lifted degeneracy in the entanglement spectrum of the SP state as
shown in Fig.~\ref{fig:supp7}(a). The quasi-degeneracy in the lowest six
levels of the entanglement spectrum for odd $l_c$ suggests dimerization in
the SP state, which is consistent with results in Figs.~\ref{fig:supp5} and %
\ref{fig:supp6}. Importantly, we note that there is no such dimerization in
the SPT state, which highlights the difference between these phases and
justifies our approach to determine the topological phase boundary from
entanglement-spectrum calculations based on bipartition.

\end{widetext}


\begin{thebibliography}{99}
\bibitem{guwen2009} Z.-C. Gu and X.-G. Wen, Phys. Rev. B {\bf 80}, 155131 (2009).

\bibitem{pollmann2010} F. Pollmann, E. Berg, A. M. Turner, and M. Oshikawa, Phys. Rev. B {\bf 85}, 075125 (2012).

\bibitem{wen89} X. G. Wen, Phys. Rev. B {\bf 40}, 7387 (1989).

\bibitem{WenNiu90}  X. G. Wen and Q. Niu, Phys. Rev. B {\bf 41}, 9377 (1990).

\bibitem{Wen90} X. G. Wen, Int. J. Mod. Phys. B {\bf B4}, 239 (1990).

\bibitem{Haldane83} F. D. M. Haldane, Phys. Rev. Lett. {\bf 50}, 1153 (1983); Phys. Lett. A {\bf 93}, 464 (1983).

\bibitem{hall1} C. L. Kane and E. J. Mele, Phys. Rev. Lett. {\bf 95}, 146802 (2005).

\bibitem{hall2} B. A. Bernevig and S.-C Zhang, Phys. Rev. Lett. {\bf 96}, 106802 (2006).

\bibitem{hall3} J. E. Moore and L. Balents, Phys. Rev. B {\bf 75}, 121306(R) (2007).

\bibitem{hall4} L. Fu, C. L. Kane, and E. J. Mele, Phys. Rev. Lett. {\bf 98}, 106803 (2007).

\bibitem{hall5} X.-L. Qi, T. Hughes, and S.-C. Zhang, Phys. Rev. B {\bf 78}, 195424 (2008).

\bibitem{ChenGuWen} X. Chen, Z.-C. Gu, and X.-G. Wen, Phys. Rev. B {\bf 83}, 035107 (2011).

\bibitem{ChenGuLiuWen} X. Chen, Z.-C. Gu, Z.-X. Liu, and X.-G. Wen, Phys. Rev. B {\bf 87}, 155114 (2013).

\bibitem{Kitaev} A. Kitaev, {\it the Proceedings of the L. D. Landau Memorial Conference  "Advances in Theoretical Physics", June 22-26, 2008, Chernogolovka, Moscow region, Russia}

\bibitem{Shinsy} S. Ryu, A. Schnyder, A. Furusaki, and A. Ludwig, New J. Phys. {\bf 12}, 065010 (2010).


\bibitem{Kitaev2} L. Fidkowski and A. Kitaev, Phys. Rev. B {\bf 81}, 134509 (2010).

\bibitem{GuWen} Z.-C. Gu and X.-G. Wen, Phys. Rev. B {\bf 90}, 115141 (2014).

\bibitem{WangSenthil} C. Wang, A. C. Potter, and T. Senthil, Science {\bf 343}, 629 (2014).

\bibitem{HQWu} H.-Q. Wu, Y.-Y. He, Y.-Z. You, T. Yoshida, N. Kawakami, C. Xu, Z. Y. Meng, and Z.-Y. Lu, Phys. Rev. B {\bf 94}, 165121 (2016).


\bibitem{AE1}  M. Takamoto, F. L. Hong, R. Higashi, and H. Katori, Nature (London) {\bf 435}, 321 (2005).

\bibitem{AE2} A. D. Ludlow, M. M. Boyd, T. Zelevinsky, S. M. Foreman, S. Blatt, M. Notcutt, T. Ido, and J. Ye, Phys. Rev. Lett. {\bf 96}, 033003 (2006).

\bibitem{AEnew7} M. D. Swallows, M. Bishof, Y. Lin, S. Blatt, M. J. Martin, A. M. Rey, and J. Ye, Science {\bf 331}, 1043 (2011).

\bibitem{AE3} B. J. Bloom, T. L. Nicholson, J. R. Williams, S. L. Campbell, M. Bishof, X. Zhang, W. Zhang, S. L. Bromley, and J. Ye, Nature (London) {\bf 506}, 71 (2014).

\bibitem{AEnew3} M. A. Cazalilla and A. M. Rey, Rep. Prog. Phys. {\bf 77}, 124401 (2014).

\bibitem{AE4} A. V. Gorshkov, A. M. Rey, A. J. Daley, M. M. Boyd, J. Ye, P. Zoller, and M. D. Lukin, Phys. Rev. Lett. {\bf 102}, 110503 (2009).

\bibitem{congjun03} C. Wu, J. Hu, and S. C. Zhang, Phys. Rev. Lett. {\bf 91}, 186402 (2003).

\bibitem{AE5} T. Fukuhara, Y. Takasu, M. Kumakura, and Y. Takahashi, Phys. Rev. Lett. {\bf 98}, 030401 (2007).

\bibitem{AEnew1} M. A. Cazalilla, A. F. Ho, and M. Ueda, New J. Phys. {\bf 11}, 103033 (2009).

\bibitem{AEnew4} S. Stellmer, M. K. Tey, B. Huang, R. Grimm, and F. Schreck, Phys. Rev. Lett. {\bf 103}, 200401 (2009).

\bibitem{AEnew5} B. J. DeSalvo, M. Yan, P. G. Mickelson, Y. N. Martinez de Escobar, and T. C. Killian, Phys. Rev. Lett. {\bf 105}, 030402 (2010).

\bibitem{AEnew2} A. V. Gorshkov, M. Hermele, V. Gurarie, C. Xu, P. S. Julienne, J. Ye, P. Zoller, E. Demler, M. D. Lukin, and A. M. Rey, Nat. Phys. {\bf 6}, 289 (2010).

\bibitem{phasePRL} K. Kobayashi, M. Okumura, Y. Ota, S. Yamada, and M. Machida, Phys. Rev. Lett. {\bf 109}, 235302 (2012).

\bibitem{phaseEPL} H. Nonne, M. Moliner, S. Capponi, P. Lecheminant, and K. Totsuka, Europhys. Lett. {\bf 102}, 37008 (2013).

\bibitem{quella13} K. Duivenvoorden and T. Quella, Phys. Rev. B {\bf 87}, 125145 (2013).

\bibitem{ofr1} X. Zhang, M. Bishof, S. L. Bromley, C. V. Kraus, M. S. Safronova, P. Zoller, A. M. Rey, and J. Ye, Science {\bf 345}, 1467 (2014).

\bibitem{ofr2} F. Scazza, C. Hofrichter, M. H\"ofer, P. C. De Groot, I. Bloch, and S. F\"olling, Nat. Phys. {\bf 10}, 779 (2014); {\it ibid} {\bf 11}, 514 (2015).

\bibitem{ofr3} G. Cappellini, M. Mancini, G. Pagano, P. Lombardi, L. Livi, M. Siciliani de Cumis, P. Cancio, M. Pizzocaro, D. Calonico, F. Levi, C. Sias, J. Catani, M. Inguscio, and L. Fallani, Phys. Rev. Lett. {\bf 113}, 120402 (2014); {\it ibid} {\bf 114}, 239903 (2015).


\bibitem{AEnew8} M. Mancini, G. Pagano, G. Cappellini, L. Livi, M. Rider, J. Catani, C. Sias, P. Zoller, M. Inguscio, M. Dalmonte, and L. Fallani, Science {\bf 349}, 1510 (2015).

\bibitem{phases1} V. Bois, S. Capponi, P. Lecheminant, M. Moliner, and K. Totsuka, Phys. Rev. B {\bf 91}, 075121 (2015).

\bibitem{quella15} A. Roy and T. Quella, arXiv:1512.05229.

\bibitem{AEnew9} C. Hofrichter, L. Riegger, F. Scazza, M. H\"ofer, D. R. Fernandes, I. Bloch, and S. F\"olling, Phys. Rev. X {\bf 6}, 021030 (2016).

\bibitem{phases2} V. Bois, P. Fromholz, and P. Lecheminant, Phys. Rev. B {\bf 93}, 134415 (2016).

\bibitem{phases3} S. Capponi, P. Lecheminant, and K. Totsuka, Ann. Phys. {\bf 367}, 50 (2016).

\bibitem{ye2016old} M. L. Wall, A. P. Koller, S. Li, X. Zhang, N. R. Cooper, J. Ye, and A. M. Rey, Phys. Rev. Lett. {\bf 116}, 035301 (2016).

\bibitem{ye2016} S. Kolkowitz, S. L. Bromley, T. bothwell, M. L. Wall, G. E. Marti, A. P. Koller, X. Zhang, A. M. Rey, and J. Ye, Nature (London) {\bf 542}, 66 (2017).

\bibitem{fallani2016} L. F. Livi, G. Cappellini, M. Diem, L. Franchi, C. Clivati, M. Frittelli, F. Levi, D. Calonico, J. Catani, M. Inguscio, and L. Fallani, Phys. Rev. Lett. {\bf 117}, 220401 (2016).

\bibitem{gyuboong} B. Song, C. He, S. Zhang, E. Hajiyev, W. Huang, X.-J. Liu, and G.-B. Jo, Phys. Rev. A {\bf 94}, 061604 (2016).

\bibitem{ren1} R. Zhang, Y. Cheng, H. Zhai, and P. Zhang, Phys. Rev. Lett. {\bf 115}, 135301 (2015).

\bibitem{ofrexp1} G. Pagano, M. Mancini, G. Cappellini, L. Livi, C. Sias, J. Catani, M. Inguscio, and L. Fallani, Phys. Rev. Lett. {\bf 115}, 265301 (2015).

\bibitem{ofrexp2} M. H\"ofer, L. Riegger, F. Scazza, C. Hofrichter, D. R. Fernandes, M. M. Parish, J. Levinsen, I. Bloch, and S. F\"olling, Phys. Rev. Lett. {\bf 115}, 265302 (2015).

\bibitem{magic} V. A. Dzuba and A. Derevianko, J. Phys. B {\bf 43}, 074011 (2010).

\bibitem{zeemanshift1} S. G. Porsev, A. Derevianko, and E. N. Fortson, Phys. Rev. A {\bf 69}, 021403 (2004).

\bibitem{zeemanshift2} S. G. Porsev and A. Derevianko, Phys. Rev. A {\bf 69}, 042506 (2004).

\bibitem{highband1} J.-S. Pan, X.-J. Liu, W. Zhang, W. Yi, and G.-C. Guo, Phys. Rev. Lett. {\bf 115}, 045303 (2015).

\bibitem{highband2} L. Zhou and X. Cui, Phys. Rev. B {\bf 92}, 140502(R) (2015).

\bibitem{highband3} J.-S. Pan, X.-J. Liu, W. Zhang, W. Yi, and G.-C. Guo, arXiv:1509.02993.

\bibitem{highband4} J.-S. Pan, W. Zhang, W. Yi, and G.-C. Guo, Phys. Rev. A {\bf 94}, 043619 (2016).


\bibitem{ren2} R. Zhang, D. Zhang, Y. Cheng, W. Chen, P. Zhang, and H. Zhai, Phys. Rev. A {\bf 93}, 043601 (2016).

\bibitem{cirfr} T. Bergeman, M. G. Moore, and M. Olshanii, Phys. Rev. Lett. {\bf 91}, 163201 (2003).

\bibitem{supp} See Supplemental Materials, which includes Ref.~\cite{suppref1}, for details on the topological edge modes, the characterization of the topological phase transitions, the topological invariant of the interacting fermionic SPT state, and the impact of spin-orbit coupling.

\bibitem{suppref1} X. Chen, Z.-C. Gu, and X.-G. Wen, Phys. Rev. B {\bf 84}, 235128 (2011).

\bibitem{liu2013} X.-J. Liu, Z.-X. Liu, and M. Cheng, Phys. Rev. Lett. \textbf{110}, 076401 (2013).

\bibitem{tangwen} E. Tang and X.-G. Wen, Phys. Rev.Lett. {\bf 109}, 096403 (2012).

\bibitem{MFM} T. Morimoto, A. Furusaki, and C. Mudry, Phys. Rev. B {\bf 92}, 125104 (2015).

\bibitem{dmrg1} S. R. White, Phys. Rev. Lett. {\bf 69}, 2863 (1992).

\bibitem{dmrg2} U. Schollw\"{o}k, Rev. Mod. Phys. {\bf 77}, 259 (2005).

\bibitem{Zhao2015} J.-Z. Zhao, S.-J Hu, and P. Zhang, Phys. Rev. Lett. \textbf{115}, 195302 (2015).

\bibitem{Yoshida2014} T. Yoshida, R. Peters, S. Fujimoto, and N. Kawakami, Phys.
Rev. Lett. \textbf{112}, 196404 (2014).

\bibitem{Turner2011} A. M. Turner, F. Pollmann, and E. Berg, Phys.
Rev. B \textbf{83}, 075102 (2011).

\bibitem{Pollmann2010} F. Pollmann, A. M. Turner, E. Berg, and M.
Oshikawa, Phys. Rev. B \textbf{81}, 064439 (2010).

\bibitem{Fidkowski2010} L. Fidkowski, Phys. Rev. Lett. \textbf{104}, 130502 (2010).

\bibitem{Flammia2009} S. T. Flammia, A. Hamma, T. L. Hughes, and X.-G. Wen, Phys. Rev. Lett. \textbf{103}, 261601 (2009).

\bibitem{Li2008} H. Li and F. D. M. Haldane, Phys. Rev. Lett.
\textbf{101}, 010504 (2008).


\bibitem{Hastings2010} M. B. Hastings, I. Gonz\'{a}lez, A. B. Kallin, and R.
G. Melko, Phys. Rev. Lett. \textbf{104}, 157201 (2010).

\bibitem{Daley2012} A. J. Daley, H. Pichler, J. Schachenmayer, and P.
Zoller, Phys. Rev. Lett. \textbf{109}, 020505 (2012).

\bibitem{Abanin2012} D. A. Abanin and E. Demler, Phys. Rev.
Lett. \textbf{109}, 020504 (2012).

\bibitem{jiang2012} H.-C. Jiang, Z.-H. Wang, and L. Balents, Nat. Phys. \textbf{8}, 902 (2012).

\bibitem{Islam2015} R. Islam, R. Ma, P. M. Preiss, M. E. Tai, A. Lukin, M.
Rispoli, and M. Greiner, Nature (London) \textbf{528}, 77 (2015).

\bibitem{centralcharge1} P. Calabrese and J. Cardy, J. Stat. Mech. {\bf 0406} 06002 (2004).

\bibitem{centralcharge2} A. E. B. Nielsen, G. Sierra, and J. I. Cirac, Phys. Rev. A {\bf 83}, 053807 (2011).


\bibitem{ladderspt} Z.-X. Liu, Z.-B. Yang, Y.-J. Han, W. Yi, and X.-G. Wen, Phys. Rev. B {\bf 86}, 195122 (2012).

\bibitem{quella12} K. Duivenvoorden and T. Quella, Phys. Rev. B {\bf 86}, 235142 (2012).

\bibitem{blochstringorder}M. Endres, M. Cheneau, T. Fukuhara, C. Weitenberg, P. Schauss, C. Gross, L. Mazza, M. C. Banuls, L. Pollet, I. Bloch, and S. Kuhr, Science {\bf 334}, 200 (2011).

\end{thebibliography}

\begin{thebibliography}{9}
\bibitem{chenguwen} X. Chen, Z.-C. Gu, and X.-G. Wen, Phys. Rev. B \textbf{84%
}, 235128 (2011).

\bibitem{tangwen} E. Tang and X.-G. Wen, Phys. Rev. Lett. \textbf{109},
096403 (2012).

\bibitem{MFM} T. Morimoto, A. Furusaki, and C. Mudry, Phys. Rev. B \textbf{92%
}, 125104 (2015).

\bibitem{phases1} V. Bois, S. Capponi, P. Lecheminant, M. Moliner, and K.
Totsuka, Phys. Rev. B \textbf{91}, 075121 (2015).

\bibitem{annphys} S. Capponi, P. Lecheminant, and K. Totsuka, Ann. Phys.
\textbf{367}, 50 (2016).
\end{thebibliography}
\end{document}